\documentclass[aps,prl,twocolumn,groupedaddress]{revtex4}
\usepackage{graphicx,epsfig,subfigure,afterpage,bm}

\begin{document}


\title{Microchannel flow of a shear-banding fluid: enhanced confinement effect and interfacial instability}


\author{P. Nghe$^1$, S. M. Fielding$^2$, P. Tabeling$^1$, A. Ajdari$^3$}
\affiliation{$^1$Laboratoire Microfluidique, MEMS et Nanostructures, UMR Gulliver CNRS-ESPCI 7083}
\affiliation{$^2$Department of Physics, University of Durham, Science Laboratories, South Road, Durham. DH1 3LE}
\affiliation{$^3$Laboratoire Physico-Chimie Theorique, UMR Gulliver CNRS-ESPCI 7083}


\date{\today}

\begin{abstract}
Using a micro particle imaging velocity technique, we resolve for the 
first time the three dimensionnal structure of wormlike shear banding 
flows in straight microchannels. The study revealed two effects, which 
should be generic for shear banding flows: the first is a strong 
amplification of the confinement induced by the edge of the channel, the 
second is an instability of the interface between the shear bands. A 
detailed quantitative comparison of our experimental measurements with a 
theoretical study of the diffusive Johnson Segalman model leads to 
excellent agreement. Our study clarifies the nature of shear banding 
flow instabilities, and shows that, despite the challenging complexity 
of the situation and the uncertainty regarding their molecular 
structure, shear banding flows in confined geometries are amenable to 
quantitative modelling, a feature that opens pathways to their practical 
utilization.
\end{abstract}

\pacs{}

\maketitle


In many complex fluids, a strongly nonlinear coupling between flow and
microstructure can trigger the onset of flow instabilities, leading to the
formation of coexisting ``shear bands'' that support differing shear rates at
a common shear stress.  This model situation for out-of-equilibrium phase separation has been the focus of intense research
in recent years~\cite{Olmsted2008}, with wormlike micellar
surfactants the most widely studied systems~\cite{Cates2006}, both as model
materials and for their widespread application in, {\it e.g.}, oil recovery
and personal care products. Significant progress has been achieved by imaging
experimentally the local flow profile, first by
NMR~\cite{Britton1997} and later by DLS~\cite{Salmon2003a},
ultrasound~\cite{Manneville2004} or particle image
velocimetry~\cite{Masselon2008}. In this Letter we report a significant advance in such efforts,
with the first full three dimensional (3D) velocity imaging of a shear-banding
fluid. We perform this in the context of pressure driven flow in a rectilinear
microchannel of rectangular cross section.

Using this imaging, we report two flow phenomena. The first is a
strongly enhanced confinement effect induced by the edges, that we believe may prove generic in a
broad class of heterogeneous confined flows. The second is an
instability of the interface between the shear bands, characterized by
a downstream evolution from (near the inlet) a smoothly bowed
interface between the bands, to (far downstream) a modulated interface
with wavevector in the vorticity direction. Throughout, we compare
our data with a theoretical study of the diffusive Johnson-Segalman
(DJS) model in the same geometry. The excellent quantitative
agreement we find without fitting assesses the robustness of our observations in the general context of shear-banded flows. It also suggests that the behavior of wormlike micelles in complex geometries is amenable to theoretical description at a quantitative level.

In particular, the interfacial instability we observe relates to a growing body of experiments
showing complex spatio-temporal
patterns in shear-banded flow, most of which  have been to date in curved
Couette geometry. While observations of instabilities are now quite widespread in that geometry
via 1D measurements, the only previous explicit observation of a 2D
modulation of the interface is in Ref.\cite{Lerouge2006}. The theoretical understanding of this instability remains
preliminary~\cite{Cates2006}, but suggestions about its origin include (i) a bulk viscoelastic
Taylor-Couette-like instability originating in the strongly sheared
band~\cite{LARSON1990}; and (ii) an interfacial instability triggered
by the normal stress jump between the
bands~\cite{HINCH1992,Fielding2005}. An important additional
contribution of this Letter is to give evidence favouring (ii), by
eliminating the cell curvature needed for (i).

\begin{figure}
\includegraphics[scale=0.55]{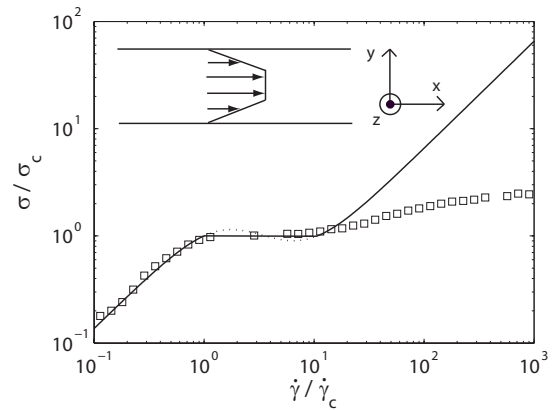}
\caption{Squares: experimental steady state flow curve \cite{Nghe2008}. Line:
  theoretical curve with the DJS model for $a=0.3, \eta=0.05$. Both are renormalized by the shear rate $\dot{\gamma}_c$ at
  the onset of shear banding, and by the value of the stress plateau
  $\sigma_c$. Correspondence is not expected at high shear rates (see
  main).  Inset: sketch of the basic (1D) shear-banded flow, with the
  coordinate system used throughout the paper.
\label{fig:rheo}}
\end{figure}


{\em Experimental ---} The sample comprises CTAB 0.3M, NaNO$_3$ 0.405M
(Acros organics), which forms a semi-dilute solution of wormlike
micelles.  Its basic rheology was characterized in
Ref.~\cite{Nghe2008}, with a stress plateau of $105 \: Pa$ beginning at
$\dot{\gamma}_c=5 \: s^{-1}$, corresponding to a pressure gradient for
the onset of shear-banding of $G_b=3.10^2\, Pa.m^{-1}$. Our experimental setup
comprises pressure driven flow in a microchannel of rectangular cross
section, with dimensions $(L_x,L_y,L_z)=(5\;cm,64\mu m,1\;mm)$ with
the axes of Fig.~\ref{fig:rheo}. The photocurable glue used in the
fabrication permits high aspect ratio $(1:16)$ channels that do not
deform even under pressures far in excess of those in this
study~\cite{Bartolo2008}. In this work, the setup of~\cite{Nghe2008}
has been substantially modified to allow full 3D velocity imaging, as
follows.  Fluorescent tracers of diameter $500\; nm$ are tracked at 50
Hz with a charged coupled device camera (Allied) through a 100x oil
immersion objective mounted on a piezo, giving a $1 \: \mu m$ thick
focal plane that can be displaced in $y$ with $10\: nm$
resolution. Images comprise $(\Delta x, \Delta z)=(36 \mu m, 72 \mu
m)$ windows, the position of which is determined using a moving table
(Marzhauser). These images can be divided into subimages along $z$,
resulting in correlations between images of size $9 \;\mu m$ for
$z$. As a whole, then, velocity is mapped over the sample with
spatio-temporal resolution $(\Delta x, \Delta y, \Delta z, \Delta
t)=(72 \: \mu m, 1 \: \mu m,9 \: \mu m, 40 \: ms)$.

{\em Theoretical ---} We compare all our data with the predictions of
the diffusive Johnson Segalman (DJS) model
\cite{JOHNSON1977}. This is not intended to give a
microscopically faithful description of our particular fluid, but is
instead the minimal phenomenological tensorial model to capture the
constitutive non-monotonicity needed for banding.  Its parameters are:
solvent viscosity $\eta$; slip parameter $a$; linear modulus $G_0$;
relaxation time $\tau$; and interfacial thickness normalized by the height $l/L_y$. Values for
these are guided by previous linear and nonlinear bulk rheological
measurements on this system, independent of the confinement effect or
instability studied here.  Specifically, the non-linear flow curve of Fig.~\ref{fig:rheo} suggests a ratio $\eta/G_0\tau \approx
0.05$ of high and low shear phase viscosities. Perfect correspondence is
impossible because the DJS model is oversimplified in having a
Newtonian high shear branch, compared to the sublinear experimental
trend. For pressure drops quite close to the onset of banding, as
here, we do not expect this limitation to be severe. We set $a=0.3$,
though our results are robust to variations in this quantity. We
estimate $l\approx(k_{\rm B}T/G_0^{exp})^{1/3}=O(10^{-8}m)$, based on linear
rheology.  The experimental aspect ratio is $L_z/L_y=16$; numerically we
explore $8, 16, 32, 64$. Numerical calculations are peformed initially in units in which $G_0=1$, 
$\tau=1$, $L_y=1$, but results are usually presented by scaling stresses (or 
pressure drops) and shear rates by their values at the onset of banding, 
as in Fig. 1. Below we will demonstrate excellent
quantitative agreement between theory and experiment for both the confinement
effect and the interfacial instability, without any fitting specific
to these phenomena.

Our numerical study considers flow driven along a channel by a
constant pressure drop $G=-\partial_x p$, assuming translational
invariance of the velocity field in the main flow direction $x$. It comprises two separate
parts, to allow a cleaner understanding of the two flow phenomena seen
experimentally.  First we study the enhanced confinement effect by simulating
channels of high aspect ratio with closed walls in $y$ and $z$. In
order to isolate this confinement effect from the interfacial instability, in
this part of the study we disallow any secondary flows in the channel
cross section, thus artificially switching off the instability. Our
aim is thereby to capture the ``basic state'' seen experimentally
close to the inlet, before the instability sets in downstream. The
second part of our study focuses on the interfacial instability, which
we now isolate from the confinement effect by switching off the latter via
the use of periodic boundaries in $z$.  This work is thereby intended
to approximate the flow seen experimentally in the central region of
$z$, well away from the lateral walls. In each run we evolve the code
to steady state $t \to \infty$ with the aim of capturing the flow pattern far
downstream from the inlet $x \to \infty$, assuming that these limits correspond to each other.  
A stringent check of our code, adapted from a former
study~\cite{Fielding2007}, is provided by comparing against a true
linear stability analysis its linear regime growth dynamics of tiny
amplitude perturbations to an initially flat interface.


\begin{figure}
\includegraphics[scale=0.48]{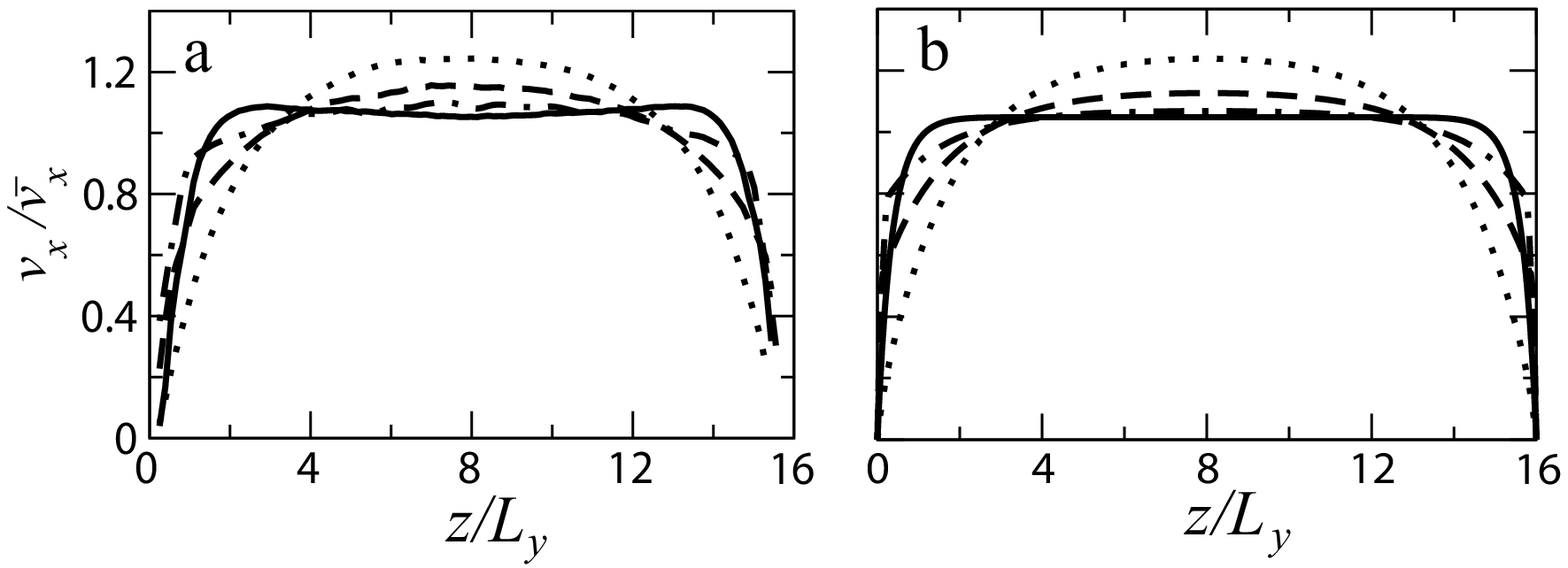}
\includegraphics[scale=0.34]{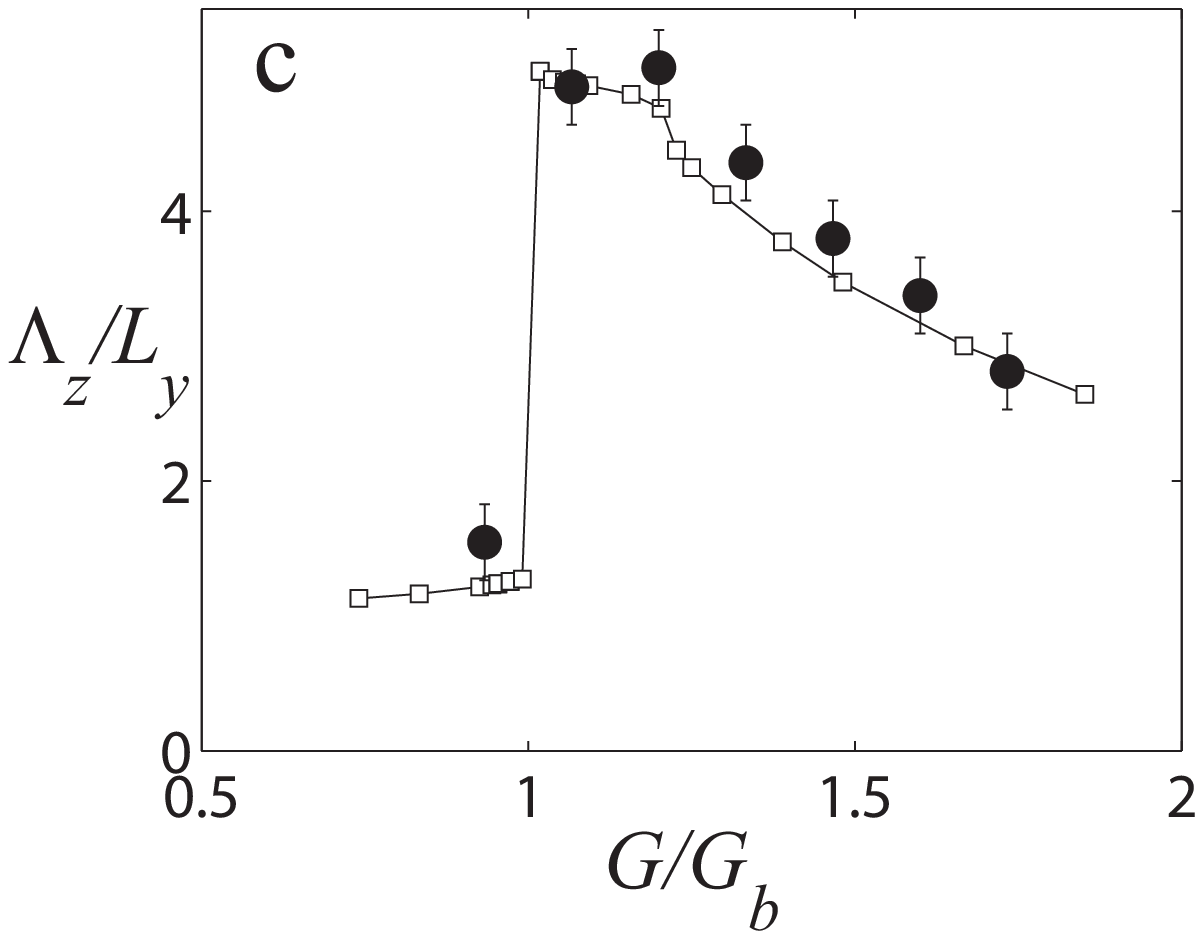}
\includegraphics[scale=0.46]{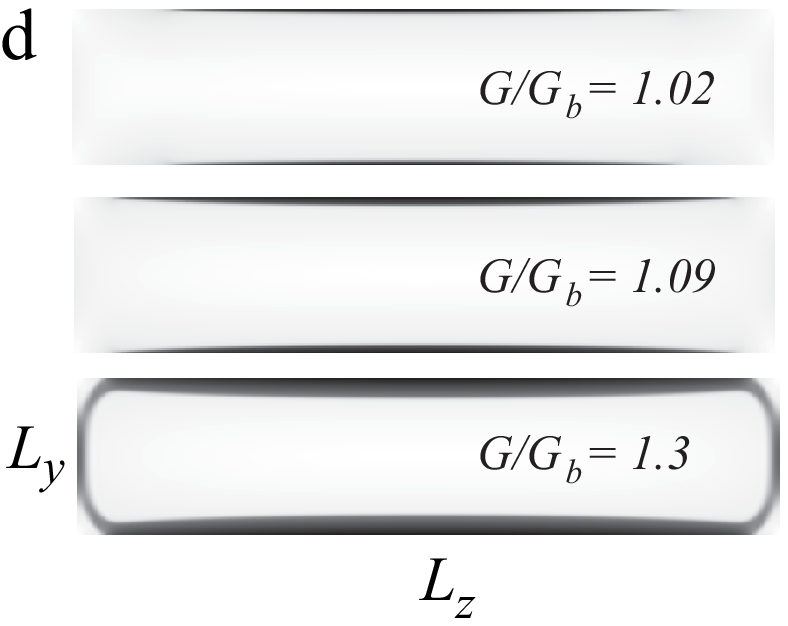}
\caption{(a) Experimental velocity profiles across the centreline
  $y=L_y/2$ of the cell cross section in the zone near the inlet (a
  typical error bar of 5\% is not shown for clarity). Pressure drop
  $G/G_{\rm b}=0.9, 1.06, 1.3, 1.6$ (solid, dotted, dashed and
  dot-dashed lines) with $G_b=3.10^2 \: Pa.m^{-1}$; (b) corresponding numerical results for the same
  scaled pressure drops with $a=0.3$, $\eta=0.05$, $l=0.01$. (c)
  Experimental (circles) and theoretical (squares) penetration length
  versus scaled pressure drop. (d) Steady state greyscale snapshot of
  invariant shear rate $|\nabla v_x|$ for pressure drop $G/G_{\rm
    b}=1.02, 1.09, 1.30$ for $L_z/L_y=8.0$ (aspect ratio rescaled for clarity).}
  \label{fig:figure2}
\end{figure}

\begin{figure}
\includegraphics[scale=0.8]{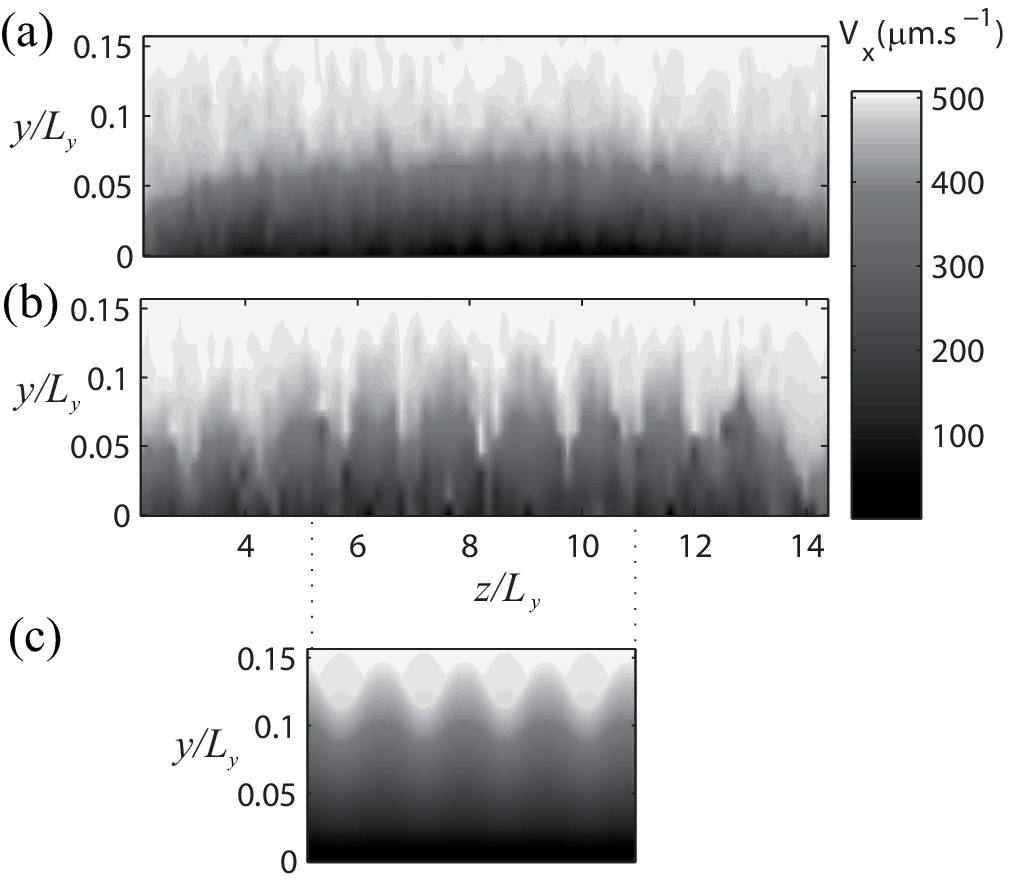}
\includegraphics[scale=0.35]{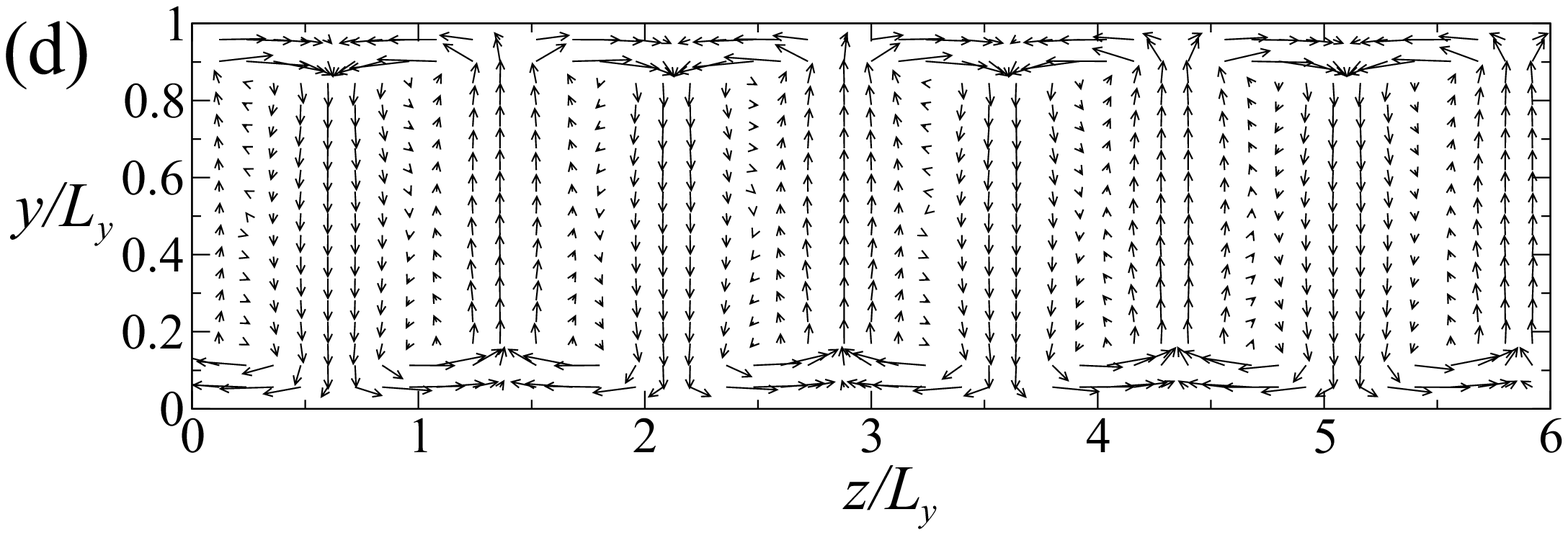}
\caption{Greyscale of the experimentally measured
  velocity component $v_x$ in the channel cross-section for $G/G_{\rm b}=1.3$, at an early
  downstream location $x=1 \; cm$ before the instability has set in (a)
  and far downstream $x= 4 \; cm$ (b), with the fully developed pattern. (c) Corresponding numerical snapshot of the
  fully developed pattern with $G/G_{\rm b}=1.3$, $a=0.3$, $\eta=0.05,
  l=0.0015, L_z=6.0$ and periodic boundaries in $z$. (d) Numerically computed secondary flow field for the same
  conditions.\label{fig:figure3}}
\end{figure}

{\em Results I: basic state and confinement effect ---} First we explore the
``basic state'' that is observed across the channel cross-section in the zone far enough from the inlet
for the local rheology to have attained a stationary banded state, but
close enough that the interfacial instability has not set in, paying
close attention to the role of the lateral walls of the channel ($z=\{0,L_z\}$) in
breaking translational invariance in the $z$ direction. 
Given the high aspect ratio of our channel ($L_z/L_y = 16$), we might expect the flow profile to be almost $z$ independant except for regions close to the lateral walls. Indeed for Newtonian flows it is known that for $L_z\gg L_y$ these regions have a size of order $L_y$. The new phenomenon we report is that for pressure gradients $G$ beyond the onset of shear-banding, the velocity field can vary significantly with $z$ across the entire channel cross-section.

To quantify this, we measure for each $G$ the centerline velocity profiles $v_c(z)=v_x(y=L_y/2,z)$  and a penetration length $\Lambda_z$, characterising the size of the lateral regions affected by the walls. $\Lambda_z$ is defined by $v_c(\Lambda_z)=0.96 \: v_m$ with $v_m=v_c(z=L_z/2)$ the maximum velocity. This criterion is chosen such that $\Lambda_z= L_y$ for Newtonian flow for $L_z \gg L_y$.
For our shear banding fluid, $v_c(z)$ profiles (Fig.~\ref{fig:figure2}a,b) and $\Lambda_z$ plotted {\it vs.} pressure
gradient (Fig.~\ref{fig:figure2}c) compare with striking quantitative
agreement between theory and experiment. 

Three regimes are
evident. Below the onset of banding (regime I, $G/G_{\rm b}< 1$) the
flow is Newtonian to good approximation: $v_c(z)$ profiles are invariant over a large zone around the centrepoints $L_z/2$ (Fig.~\ref{fig:figure2}a,b solid line) and $\Lambda_z\approx L_y$ as remarked above.
At the onset $G=G_{\rm b}$ of regime II a thin band of
high shear rate nucleates along each of the long walls $y=\{0,L_y\}$
around the centrepoints $z=L_z/2$, these being the locations at which
the shear stress is maximum and so first exceeds the plateau (banding)
stress of the bulk flow curve as $G$ is increased. These bands are
seen in our numerical $y-z$ maps of the invariant shear rate $|\nabla
v_x|$ in Fig.~\ref{fig:figure2}d, top.  It is at this onset that the
centreline profile suddenly becomes strongly bowed
(Fig.~\ref{fig:figure2}a,b, dotted line). Reflecting this is the
dramatic jump of $\Lambda_z$ to $\Lambda_z\approx 5 \: L_y$. As $G$
increases further the thickness of the high shear bands increases, and
so does their lateral spread along the long walls
(Fig.~\ref{fig:figure2}d, middle). At the onset of regime III, shear
bands finally form along the short walls (Fig.~\ref{fig:figure2}d,
bottom; and Fig.~\ref{fig:figure2}a,b long dashed and dot-dashed).
$\Lambda_z$ then shows a downward kink into a decline to
$\Lambda_z\approx L_y$ at high $G$.

How can the large value of $\Lambda_z$ in regime II be understood?
For a shear-banding fluid, smooth stress inhomogeneities across the
channel cross section are strongly magnified in the velocity field via
the nucleation of lubricating bands at the long walls. For a complete slip at these walls the system would ``forget'' the
short dimension altogether and show parabolic Poiseuille flow across
the long centreline. Simulations at fixed $L_z$ in regime II indeed
show $\Lambda_z$ to increase with decreasing viscosity of the high
shear band (not shown). This has important implications for a broad
class of flows with stress dependent wall slip, to be explored in
future.

\begin{figure}
\includegraphics[scale=0.55]{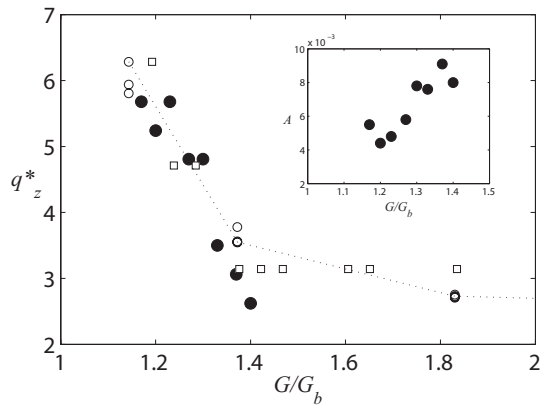}
\caption{Solid circles: experimental wavevector of the interfacial
  undulations as a function of applied pressure drop. Open squares:
  corresponding numerical results for the fully non-linear state
  computed with $\eta=0.05, l=0.0015, L_z=4.0$. The numerical
  wavevector is constrained by the cell length to be heavily quantised
  as $q_z=2n\pi/L_z$. We therefore also show results (open circles)
  for the maximally unstable mode in a linear stability analysis,
  which does not suffer quantisation. Insert shows the experimental
  amplitude $A$ in units of $L_y$.
  \label{fig:figure4}}
\end{figure}


{\em Results II: instability ---}We describe first the steady state experimental velocity profiles
$v_x(y,z)$ at different downstream locations $x$, for a fixed pressure
drop, in a cross sectional area $[z=0-1 \: mm]$ by $[y=0-14 \: \mu m]$
containing the high shear phase together with the interface near the
bottom wall.  Close to the inlet ($x=1 \: cm$, Fig.~\ref{fig:figure3}a) the interface between the bands is smoothly bowed, corresponding
to the basic state discussed above. In contrast, far downstream ($x=4
\: cm$, Fig.~\ref{fig:figure3}b) the interface has undulations
with wavevector along the $z$ axis, with wavelength of the order of
the channel height $L_y$. This pattern is steady in time, and
invariant with respect to further progression downstream. (At
intermediate downstream locations $1 \: cm <x<4 \: cm$, the
instability develops via interfacial corrugations that emerge close to
the lateral walls, and progressively invade towards the cell centre
with increasing $x$; not shown.)   

Such interface instabilities have already been predicted ~\cite{Fielding2005,Fielding2007} and observed developing in time ~\cite{Lerouge2006} in Couette geometry. Similarly to these former predictions, ongoing linear stability analysis of Poiseuille flow shows the concurrence of
unstable $q_x$ and $q_z$ interfacial modes of wavelengths comparable to $L_y$, over a range of parameter values. Unfortunately our
measurement averages velocities over $\Delta x = 72 \: \mu m > L_y$ which precludes observation of wavevectors $q_x$. We do, though, concentrate on exploring with our setup any instability to modulation in the vorticity direction. Thanks to our flow cell aspect ratio $L_z/L_y = 16$, there is a domain of $z$ around the middle of the channel ($z=L_z/2$) over which the curvature of the basic state due to the enhanced confinement effect (Fig.~\ref{fig:figure3}a) is negligible compared to the interface modulation for around ten wavelengths (Fig.~\ref{fig:figure3}b). We perform our analysis in this large aspect ratio domain that can be considered invariant in $z$. Our corresponding numerics are therefore performed with periodic boundaries in $z$, assuming translational invariance in $x$. The numerical counterpart of the fully developed velocity pattern shows good correspondence (Fig.~\ref{fig:figure3}c).

Associated with the interfacial undulations are counter-rotatory
streamwise vortices seen numerically in the secondary flow field $v_y,
y_z$ (Fig.~\ref{fig:figure3}d). Experimentally, focal plane
constraints make measurement of $v_y$, and so direct comparison with this 
numerical map impossible. However an analysis of $v_z$ (not shown)
indicates spatial modulation at an identical wavelength to that in
$v_x$, at least consistent with streamwise recirculation.

We analyse the developed pattern by measuring the wavelength $2\pi/q_z$ of the
interfacial undulations across a range of $G$. We compare the size of the wavevector $q_z^{\star}$ with
linear stability analysis and full non-linear simulations of the final
steady state in the DJS model (Fig. 4). For the range explored we find
a decreasing wavevector, and good quantitative agreement between
theory and experiment  (higher values of $G$ for which saturation
of wavevector is predicted numerically are unattainable
in our experiments). The amplitude $A$ of the interface position undulations in $y$ (Fig.4, inset)
increases with $G$ and is a few percent of the channel height.

{\em Conclusion ---} Careful 3D velocimetric study of a
shear banding wormlike micellar surfactant solution in microchannel
flow reveals a strongly enhanced confinement effect induced by the edges, and an interfacial
instability with wavevector in the vorticity direction. Striking
quantitative agreement is obtained with a numerical study of the DJS
model, without free parameter fitting.  Given that this model is
highly phenomenological, with only crude representation of the
microscopic rheology, this suggests robustness of the phenomena to underlying microscopic details, hinging only
on the presence of a non-monotonic constitutive curve and a normal
stress jump across the interface between the shear bands. The confinement effect, here due to inhomogeneous banding on the walls, may prove generic in a broader class of inhomogeneous confined flows, a conjecture we shall explore more fully in future work.

Our observation of the interfacial instability in planar flow, combined with its linearly unstable nature, provides strong evidence in favour of scenario (ii) of the introduction. Experimental resolution unfortunately averages possible $q_x$ modes, predicted by linear stability analysis to be concurrent with $q_z$ ones (ongoing work). It thus remains an open
challenge both theoretically and experimentally to delineate the
effect of nonlinear coupling between these modes; and to study the
interplay between the interfacial instability and the enhanced confinement
effect reported here.

SMF thanks Mike Evans and Peter Olmsted for discussions; and the UK's EPSRC for funding
(EP/E5336X/1). The authors acknowledge Guillaume Degr\'e, Jonathan Rothstein, Alexander Morozov, Elie Rapha\"el and Mike Cates for discussions; and Total, CNRS and ESPCI for their support.

\end{document}